\documentclass[floatfix,prb,epsf,showpacs,twocolumn,superscriptaddress]{revtex4}
\usepackage{amsfonts}
\usepackage{amsmath,amssymb,natbib,bm,graphicx,url,epsfig}
\usepackage{bm}
\usepackage[ansinew]{inputenc} 
\usepackage[active]{srcltx}
\usepackage{color}
\newcommand{\be}{\begin{equation}}
\newcommand{\ee}{\end{equation}}
\newcommand{\bes}{\begin{equation}\begin{split}}
\newcommand{\ees}{\end{split}\end{equation}}

\begin{document}

\title{ Superbosonization in disorder and chaos: The role of anomalies }
\author{Tigran A. Sedrakyan}
\affiliation{Department of Physics, University of Massachusetts Amherst, Amherst,
Massachusetts 01003, USA}
\author{Konstantin B. Efetov}
\affiliation{Theoretische Physik III, Ruhr-Universität Bochum, D-44780 Bochum, Germany}
\affiliation{National University of Science and Technology \textquotedblleft
MISiS\textquotedblright , Moscow, 119049, Russia}

\begin{abstract}
Superbosonization formula aims at rigorously calculating fermionic integrals
via employing supersymmetry. We derive such a supermatrix representation of
superfield integrals and specify integration contours for the supermatrices.
{The }derivation is essentially based on the supersymmetric
generalization of the Itzikson-Zuber integral in the presence of anomalies
in the {Berezinian} and shows how an integral over supervectors is
eventually reduced to an integral over commuting variables. {The }%
approach is tested by calculating both one and two point correlation
functions in a class of random matrix models. It is argued that the approach
is capable of producing nonperturbative results in various systems with
disorder, including physics of many-body localization, and other situations
hosting localization phenomena.

%On the mathematical level,
%In particular we show how an integral over supervectors eventually
%is reduced to an integral over commuting variables.
%, and hence the name-superbosonization...
\end{abstract}

\maketitle

%Supermatrix action as a bosonization for generating functionals:
%The role of anomalies...}
%\maketitle
%\end{center}
%\vspace{5cm}

%{\bf Abstract}
%............
%\end{center}

%\newpage

\section{Introduction}

Supersymmetry\cite{berezin1,berezin2} deals with Grassmann numbers, that
were originally invented in mathematics and later used in quantum
field theory as the classical analogues of anticommuting operators. This
mathematical construction is proven to be a very useful tool for studies in
various fields of physics and in particular in models of quantum chaos,
involving random matrix theory and various models of disorder.\cite%
{Efetov1,Efetov2,VWZ}

One of the prominent methods employing supersymmetry is the non-linear
supersymmetric $\sigma $-model\cite{Efetov1,Efetov2} description of
disordered metallic conductors. According to this standard formalism,
effective field theory is described by an action with coordinate dependent
supermatrix field, $Q(\mathbf{r})$, obeying the constraint,
\begin{equation}
Q^{2}(\mathbf{r})=1.  \label{e0}
\end{equation}%
This method has a broad range of applications including study of
Anderson localization, mesoscopic fluctuations, levels statistics in a
limited volume, quantum chaos. A general form of the free energy functional 
$F$ is rather simple
\begin{equation}
F\left[ Q\right] =\frac{\pi \nu }{8}\int Str\left[ D\left( \mathbf{\nabla }%
Q\right) ^{2}+2i\left( \omega +i\delta \right) Q\left( \mathbf{r}\right) %
\right] d\mathbf{r}  \label{e1}
\end{equation}%
containing the classical diffusion coefficient $D$, the one
particle density of states $\nu ${\ and frequency }$\omega ${%
. Although the free energy }$F\left[ Q\right] ,${\ Eq. (\ref{e1}), is
written in the limit of a weak disorder, it can be used for a strongly
disordered samples replacing the gradient by finite differences.} At very
low energies the effective {free energy functional }is dominated
{in a finite volume }by the zero spatial mode, $Q(\mathbf{r})=Q_{0}$,
which is independent of $\mathbf{r}$. In {this limit the model is
especially simple containing only the second term in Eq. (\ref{e1}). The }%
spectral properties of the theory are \emph{universal} and coincide with
those of Wigner-Dyson random matrix ensembles with corresponding
symmetries\cite{Efetov1,Efetov2,VWZ,ABG,CB}.

The derivation of the $\sigma ${-model, Eq. (\ref{e1}), from
microscopic models is not exact and is based on a saddle point method
applicable at weak disorder or the large size of the matrices in the
Wigner-Dyson ensembles. At the same time, the \textquotedblleft
diffusive\textquotedblright\ }$\sigma -${model, Eq. (\ref{e1}) is not
applicable for, e.g. description of electron motion }in ballistic regime,
where characteristic spatial scales are much smaller than the mean free
path. Another important problem, that is known to be out of the reach of
nonlinear $\sigma $-model, {is random matrix models with finite range
correlations between the matrix elements that are do not belong the
Wigner-Dyson ensembles. One of the examples are} models of weakly {%
non-}diagonal matrices.\cite{OK1,OK2}. {Of course, there are many
other models that cannot be reduced to the }$\sigma $-model, Eq. (%
\ref{e1}). 

In many of those models correlation functions of interest can be
expressed from the beginning in terms of integrals over supervectors and the
problem arise due to absence of a possibility of using the saddle-point
approximation leading to Eqs. (\ref{e0}) and (\ref{e1}). Therefore, it is
natural to try to generalize the $\sigma $-{model, Eq. (\ref{e1})
to a model containing supermatrices but without the constraint, Eq. (\ref{e0}%
). In such a model, the generating functional }$Z(J)$ {would be
expressed in terms of an integral over unconstrained supermatrices}, 
and having calculated this integral, one would be able to compute
correlation functions of interest. It should be noticed here that usually
many-level or many point correlation functions are really interesting.
One-level or one-point averages (average density of states) usually do not
bring an interesting information about the systems (in the problem of
Anderson localization, the average density of states cannot help to
distinguish between the metal and insulator).

In principle, some dual representations of a generating functional,
initially given by an integral over $N\times N$ Hermitian matrices (color
space), are known as color-flavor transformations\cite{ZirnC_F}. They
transform the original integral in \textquotedblleft color
space\textquotedblright\ to an integral over certain super-manifolds, which
are acting in the dual space (flavor space). However, being
interesting on its own, this transformation has not yet evolved into a new
computational tool. 

Trying to find a new method of studying non-standard problems of the
supersymmetry method a kind of bosonization procedure to the original
fermionic functional $Z(J)$ has been suggested some time ago \cite{E, EK}.
As a result, the partition function has been represented in a supermatrix
action formulation without any constraint what soever; this approach was
claimed to be applicable to the physics of electron motion at all scales. It
seemed that the limitations due to the non-linearity of the conventional $%
\sigma $-model representation were overcame. Nevertheless the formula of
superbosonization was not well understood from the practical point of view,
namely the integration method was not specified.

More precisely, in Ref. \onlinecite{E} a new superbosonization formula
that allowed the field-integral {over supervectors }be
expressed through a supermatrix integral has been derived,
\begin{equation}
\int D\psi D{\bar{\psi}}\;F(\psi \otimes {\bar{\psi}})=\int_{{{\mathbb{H}}%
_{n}}}D{\mathcal{A}}\;\;{\text{Sdet}}{\mathcal{A}}^{-1}\;F({\mathcal{A}}).
\label{formula}
\end{equation}%
where ${\mathbb{H}_{n}}$ is the linear space of %hermitian
$n$-dimensional complex supermatrices, $\psi \in U(n,1|n,1)$ and ${\bar{\psi}%
}\in U(n,1|n,1)$ are supervectors, and $F:{\mathbb{H}}_{n}\rightarrow {%
\mathbb{G}}$ is a formal map with ${\mathbb{G}}$ representing a superspace.%
\cite{footnote} Importantly, the right-hand-side of Eq.~(\ref%
{formula}) {could} be evaluated under general conditions, without
reducing it to any mean-field manifold. For this reason, {it was
suggested that }Eq.~(\ref{formula}) {could be} capable of producing
non-perturbative results in various models of disorder. {One can
imagine that Eq. (\ref{formula})} {can represent }a promising
approach for non-perturbative studies in physics of many-body localization%
\cite{MBL,Altland} and other situations where disorder plays an important role.\cite{
general}

Originally\cite{E}, Eq. (\ref{formula}) has been derived rather
schematically without discussing contours of integration over the commuting
elements of the supermatrix ${\mathcal{A}}.$ An attempt to specify
contours of integration has been undertaken in Ref.~\onlinecite{YEK}. Roughly,
speaking it was suggested to integrate over the eigenvalues of the
boson-block from $-\infty $ {to} $\infty ,$ {while the
integration over the eigenvalues of the fermion-fermion block has to be
performed over a compact domain (a circle in the simplest case).
Surprisingly, it turned out that such an integration was well defined only
in rather uninteresting cases. In particular, it worked perfectly well for
correlation functions that required a sufficiently small number }$q\leq n$
of the bosonic components, where $n$ was a number of
artificial \textquotedblleft orbitals\textquotedblright . In other words,
one could use Eqs. (\ref{formula}) for calculation of the density of states
in case of the unitary ensemble, while one encountered a singularity of the
type $\infty \times 0,$ {when trying to calculate a two-level
correlation functions. The situation for, e.g. orthogonal ensemble was even
worse and one could not calculate even the density of states in this case.
The situation was better when using a sufficiently large number of the
\textquotedblleft orbitals\textquotedblright\ }$n$ but this could be
efficiently closer to results obtained using the standard saddle-point
method and therefore less interesting. 

These findings have been confirmed rigorously in Ref.~\onlinecite%
{Littelmann} but the case $q>n$ was not resolved and it was even
concluded that the superbosonization formula, Eq. (\ref{formula}), was not
correct for this case. This was a serious obstacle in using the
superbosonization for applications to interesting unsolved problems.

In this paper we resolve this long standing problem of the
integration in Eq. (\ref{formula}) for the case of hermitian matrices with
an arbitrary correlation between the matrix elements. Of course, the
suggested approach can be used for disordered systems with a broken
time-reversal invariance. We do it integrating over the eigenvalues of the
fermion-fermion block along the imaginary axis from $-i\infty $ {to }%
$i\infty $ {instead of the integration along the circle adopted in
Refs.~\onlinecite{YEK,Littelmann}. This does not make a difference in the results
for }$q\leq n$ {but it makes the integral, Eq. (\ref{formula}), well
defined for }$q>n$ and computation of many point correlation
functions feasible, thus establishing a new method of calculations for
interesting problems.  \ 

The paper is organized as follows. In Section II we set the basis for the
subsequent analysis of the bosonization procedure of Ref. \onlinecite{E}
by calculation of a supersymmetric generalization of Itzykson-Zuber
(IZ) integral. In Section III we show how the formulated supermatrix
representation of integrals {over supervectors }(the so called
bosonized representation) can be evaluated. In particular, we
derive the domains of integration, for which the bosonization formula is
exact. It is remarkable that this regularized scheme leads to an effective
reduction of dimensionality of the domain of integration, which is \emph{%
noncompact}.

The proof is essentially based on the results discussed in Section II: the
supersymmetric generalization of the Itzykson-Zuber (IZ) integral, \cite{IZ,
G, S, AM, GW, TS} in situations when {a} boundary term is crucial due to
the presence of singularities in the Berezinian. Emergence of this boundary
term in the IZ integral ensures that both representations of the generating
functional coincide.

%By contrast, for the same case
%%for example in the case of $4\times 4$ supermatrices,
%in traditional non-linear $\sigma$-model one has to integrate both over larger number of
%both, Grassmann and bosonic variables, while in fermionic
%representation the integral is $m^2$ dimensional.
%Another
%difference is that in our regularized scheme with effectively
%reduced dimensionality, the domain of integration over bosonic
%variables is {\em noncompact}.

%In Ref. \onlinecite{GW} it has been shown, that the standard
%result for the IZ integral, presented in \cite{G, AM} must be
%corrected by the boundary Efetov-Wigner type term, which is
%crucial in the presence of singularities.
%We show, that

In Section IV we apply the regularized superbosonization formula
to calculation of correlation functions in random matrix models. We
derive both one and two point correlation functions for Hermitian diagonal
random matrices with continuously distributed components and correction to
the density of states for weakly non-diagonal random matrices.\cite{OK1,OK2}
Technical details of some of the derivations are presented in Appendices A,
B, and C.

%\section{The formula}

\section{Anomaly in supersymmetric Itzykson-Zuber integral}

\subsection{Supersymmetric Itzykson-Zuber integral}

In this section we present useful formulae, that will be applied in
subsequent sections. Let us note that in all future considerations the
integration over the linear space of %hermitian
complex supermatrices, $\int_{\mathbb{H}_{n}}D{\mathcal{A}}$, with flat
Berezin measure\cite{berezin1,berezin2} is always performed first by
diagonalizing the matrix ${\mathcal{A}}$ and then by integrating over the
eigenvalues. {We distinguish between \textquotedblleft
fermion-fermion (FF)\textquotedblright\ and \textquotedblleft boson-boson
(BB)\textquotedblright\ blocks of the matrix }${\mathcal{A}}$ {%
corresponding respectively to products }$\psi _{F}\otimes {\bar{\psi}}_{F}$
{and} $\psi _{B}\otimes {\bar{\psi}}_{B}$ {of anticommuting
and commuting components of the supervectors. After the diagonalization of
the supermatrix }${\mathcal{A}}$ one half of the eigenvalues will be
in the FF-block, and the other part will be in the BB-block. We will call
these eigenvalues FF- and BB-eigenvalues respectively. 

We will demonstrate that the integration over the BB-eigenvalues
should be performed in the infinite interval ${{\mathbb{R}}}\equiv
\{-\infty ,\infty \}$, {while} {the} integration over {%
the FF-}eigenvalues {should be} performed in the infinite interval $%
\{-i\infty ,i\infty \}$. {This contrasts the integration rules of
Refs. \onlinecite{YEK,Littelmann}, where the integration over the FF-eigenvalues
was performed along the unit circle. }Note, that any complex $2n\times 2n$
supermatrix, ${\mathcal{A}}$, can be diagonalized as ${\mathcal{A}}=U{%
\mathcal{A}}_{d}{\bar{V}}$, where $U\in U(n\mid n)$, ${\bar{V}}\in {U(n\mid
n)}/{U^{2n}(1)}$ are diagonalization matrices restricted correspondingly to 
the unitary supergroup and its subspace with removed phases.

Here we are interested in Itzykson-Zuber integral of the type
\begin{eqnarray}
&&\Gamma \left[ \{{\tilde{b}}_{j}{,}b_{j},\}\mid \{\tilde{\lambda}%
_{j},\lambda {_{j}}\}\right] \qquad \qquad \qquad \qquad \qquad  \label{IZI}
\\
&=&\int DUD{\bar{V}}\;\exp \left\{ {\text{Re}}\;{\text{Str}}\left[ UB_{d}{%
\bar{V}}Q_{d}\right] \right\} ,  \notag
\end{eqnarray}%
where $B_{d}={\text{diag}}\{\tilde{b}_{i},b_{i}\}$ and $Q_{d}={\text{diag}}\{%
\tilde{\lambda},\lambda ,\}$ are the {FF and BB }eigenvalues {%
of }supermatrices $B$ and $Q$ respectively and $U\in U(n\mid n)$, ${\bar{V}}%
\in {U(n\mid n)}/{U^{2n}(1)}$. Then, the result of integration reads \cite%
{G, AM, GW}
\begin{eqnarray}
&&\Gamma \left[ \{{\tilde{b}_{j},}b,\}\mid \{{\tilde{\lambda}_{j},}\lambda
_{j}\}\right] =\left[ 1-\eta \left( \{{\tilde{\lambda}_{i},}\lambda
_{i}\}\right) \right]  \label{15} \\
&&\times \;\frac{\prod_{i}\delta ({\tilde{b}_{i}})\delta (b_{i})}{\Delta ^{2}%
\bigl(\{{\tilde{b}_{j}}^{2},b_{j}^{2}\}\bigr)}+\Gamma _{0}\left[ \{{\tilde{b}%
_{j},}b_{j}\}\mid \{{\tilde{\lambda}_{j},}\lambda _{j}\}\right] .  \notag
\end{eqnarray}%
Here $\Gamma _{0}\left[ \{{\tilde{b}_{j},}b_{j}\}\mid \{{\tilde{\lambda}_{j},%
}\lambda _{j}\}\right] $ is the result of the bulk integration without accounting for the singularity in the Berezinian (if there is such). It has
the form
\begin{eqnarray}
&&\!\!\!\Gamma _{0}\left[ \{{\tilde{b}_{j},}b_{j}\}\mid \{{\tilde{\lambda}%
_{j},}\lambda _{j}\}\right]  \label{NIZ} \\
&&\!\!\!=\frac{1}{{2^{2n^{2}}(n!)^{2}}}\;\frac{\det {J_{0}[{\tilde{b}_{p}}{%
\tilde{\lambda}_{q}}}]_{p,q=1\dots n}\det {J_{0}[b_{l}\lambda _{m}]}%
_{l,m=1\dots n}}{\Delta \left( \{{\tilde{b}_{j}}^{2},b_{j}^{2}\}\right)
\Delta \left( \{{\tilde{\lambda}_{j}}^{2},\lambda _{j}^{2}\}\right) },
\notag
\end{eqnarray}%
where
\begin{eqnarray}
\Delta \left( \{{\tilde{b}_{j}}^{2},b_{j}^{2}\}\right) &=&\frac{%
\prod_{k<r=1}^{n}({\tilde{b}}_{k}^{2}-{\tilde{b}}_{r}^{2})%
\prod_{l<m=1}^{n}(b_{i}^{2}-b_{m}^{2})}{\prod_{p<q=1}^{n}({\tilde{b}}%
_{q}^{2}-b_{p}^{2})}  \label{14} \nonumber\\
&=&\det \left[ \frac{1}{{\tilde{b}}_{j}^{2}-b_{i}^{2}}\right] _{i,j=1\dots n}
\end{eqnarray}%
is the supersymmetric Vandermonde determinant and $J_{0}[b_{p}\lambda _{q}]$
is the zero-order Bessel function. The term $\eta \left( \{{\tilde{\lambda}%
_{i},}\lambda _{i}\}\right) $ is the boundary term arising from the
singularities of the Berezinian (This type of the boundary term in the
integrals over supermatrices has been found in {Refs.~\onlinecite%
{level-level,Efetov1}} and is {sometimes }called Efetov-Wegner
boundary term\cite{Verbaarschot1,Kieburg}). It originates from the regularization of the anomaly in the
Berezinian and is given by~\cite{GW}
\begin{eqnarray}
\eta \left( \{{\tilde{\lambda}_{i},}\lambda _{i}\}\right) &=&\frac{1}{\Delta %
\bigl(\{{\tilde{\lambda}_{j}}^{2},\lambda _{j}^{2}\}\bigr)}\;  \label{17} \\
&\times &\det \Biggl[\frac{1}{{{\tilde{\lambda}_{l}}^{2}-\lambda _{k}^{2}}}%
\left( 1-e^{\frac{{{\tilde{\lambda}_{l}}^{2}-\lambda _{k}^{2}}}{2t}}\right) %
\Biggr]_{k,l=1,\dots ,n}.  \notag
\end{eqnarray}

One can easily check, that the expression (\ref{14}) for IZ integral  boundary terms,
$\Gamma _{0}\left[ \{{\tilde{b}_{j},}b_{j}\}\mid \{{\tilde{\lambda}_{j},}%
\lambda _{j}\}\right] $, does not fulfill {Eq.} (\ref{13}) ({%
see below).} The singularity of the Berezinian $\Delta ^{2}(\{{\tilde{b}_{j}%
}^{2},b_{j}^{2}\})$ in (\ref{13}) gives rise to the appearance of {%
the }boundary term in $\Gamma \Bigl[\{{\tilde{b}_{j},}b_{j}\}\mid \{{\tilde{%
\lambda}_{j},}\lambda _{j}\}\Bigr]$.

\subsection{Origin of the boundary term in the Itzykson-Zuber integral}

Our aim in this section is to underline the origin of the anomaly of the
Berezinian and the implication for the supersymmetric Itzykson-Zuber
integral. For this purpose for any given diagonal complex supermatrix $Q_{d}$
consider the Gaussian integral,

\begin{eqnarray}
&&\int DB\;\exp \left\{ \frac{1}{2t}\;{\text{Str}}[(B-Q_{d})^{2}]\right\} =
\label{13} \\
&=&\int \prod_{i}d{\tilde{b}_{i}}db_{i}\;\Delta ^{2}\left( \{{\tilde{b}_{j}}%
^{2},b_{j}^{2}\}\right) \;\exp \left\{ -\frac{1}{2t}\;{\text{Str}}%
[B_{d}^{2}+Q_{d}^{2}]\right\}  \notag \\
&&\times \int DU\;D{\bar{V}}\;\exp \left\{ \frac{1}{2t}\;{\text{Re}}\;{\text{%
Str}}\{UB_{d}{\bar{V}}Q_{d}\}\right\}  \notag \\
&=&\int \prod_{i}d{\tilde{b}_{i}}db_{i}\Delta ^{2}\left( \{{\tilde{b}_{j}}%
^{2},b_{j}^{2}\}\right) \;\Gamma \left[ \{{\tilde{b}_{j},}b_{j},\}\mid \{{%
\tilde{\lambda}_{j},}\lambda _{j}\}\right]  \notag
\end{eqnarray}%
where we have {the }diagonalized complex supermatrix $B\rightarrow
B_{d}=\text{diag}\{b_{i},\bar{b}_{i}\}$. In Eq. (\ref{13}) $\Delta ^{2}(\{{%
\tilde{b}_{j}}^{2},b_{j}^{2}\})$ is the Berezinian of the transformation, $%
B=UB_{d}{\bar{V}}$, where $U\in U(n\mid n)$ and ${\bar{V}}\in {U(n\mid n)}/{%
U^{2n}(1)}$.

The integral, Eq. (13), is originally gaussian and integrating
separately over all matrix elements of the supermatrix $B$ gives
unity. It is clear that changing the variables of the integration cannot
modify this result and one must obtain unity also integrating over the
eigenvalues. However, one can easily check, that the "naive" expression for
Itzykson-Zuber integral $\Gamma _{0}\left[ \{{\tilde{b}_{j},}b_{j}\}\mid \{{%
\tilde{\lambda}_{j},}\lambda _{j}\}\right] $, Eq. (\ref{NIZ}, \ref{14}),
{is not equal to unity.} It is the singularity of the Berezinian $%
\Delta ^{2}(\{{\tilde{b}_{j}}^{2},b_{j}^{2}\})$ in Eq. (\ref{13}), that
gives rise to the appearance of boundary term $\eta \left( \{{\tilde{\lambda}%
_{i},}\lambda _{i}\}\right) $ Eq. (\ref{17}), in $\Gamma \left[ \{{\tilde{b}%
_{j},}b_{j}\}\mid \{{\tilde{\lambda}_{j},}\lambda _{j}\}\right] $ , that was
found in Ref.~\onlinecite{GW}. Existence of this boundary term ensures the condition
that the integral Eq. (\ref{13}) is unity. Hence, the correct answer for
supersymmetric Itzykson-Zuber integral $\Gamma $ has the form Eq. (\ref{15}).

The following remark is in order. The result, $\Gamma _{0}\left[ \{{\tilde{b}%
_{j},}b_{j}\}\mid \{{\tilde{\lambda}_{j},}\lambda _{j}\}\right] $, of
{the }evaluation of the supersymmetric IZ integral in the absence of
singularities was derived by solving the supersymmetric heat equation\cite%
{G, AM}; technique, that was developed in Ref. \onlinecite{IZ} {for
conventional matrices}. It is straightforward to check that the boundary
term $\propto (1-\eta )$ in Eq. (\ref{15}) also satisfies the heat equation.

%\subsection{Identity II}

%\subsection{Identity III}

\section{Superbosonization: proof and integration contours}

%of the generating functional and its supermatrix formulation}

In this section we present a derivation of {the }superbosonization
formula and, in particular, {of }the bosonized $\sigma $-model for
random matrices. The derivation is similar to the procedure developed in Refs.~\onlinecite{Efetov1,Efetov2, G, VWZ}, but here instead of the Hubbard-{\text{Str}}%
atonovich transformation, which in the standard scheme follows the averaging
over random matrices, we use the identities from the above section. {%
Actually, the scheme of the derivation is very close to that of Ref.~\onlinecite{E}
but is more rigorous. } It is useful to recall, that formal sums of formal
products $\Psi \otimes {\bar{\Psi}}$, where $\Psi \in U(n,1|n,1)$ and ${\bar{%
\Psi}}\in U(n,1|n,1)$ are supervectors, constitute a vector space. This
vector space is defined, up to isomorphism, by the condition that every
antisymmetric, bilinear map $f:U(n,1|n,1)\times {\bar{U}}(n,1|n,1)%
\rightarrow {\mathbb{G}}$ determines a unique linear map $%
g:U(n,1|n,1)\otimes {\bar{U}}(n,1|n,1)\rightarrow {\mathbb{G}}$ with $f(\Psi
,{\bar{\Psi}})=g(\Psi \otimes {\bar{\Psi}})$. This implies that if we
consider a map, $F:{\mathbb{H}}_{n}\rightarrow {\mathbb{G}}$, then the
integral
\begin{equation}
I_{F}=\int D\psi D{\bar{\psi}}\;F(\psi \otimes {\bar{\psi}})  \label{int}
\end{equation}%
is now well defined. From now on we will restrict ourselves to the case of
maps, $F$, such that the integral $I_{F}$ in Eq. (\ref{int}) is convergent.

As the first step we make use of the identity derived in Appendix~A to
rewrite the field integral in the left-hand-side of Eq. (\ref{formula}) as

\begin{eqnarray}
&&I_{F}=\int_{{{\mathbb{H}}_{n}}}D{\mathcal{A}}F({\mathcal{A}})\int D\psi D{%
\bar{\psi}}\int_{{{\mathbb{H}}_{n}}}DB\;  \label{7} \\
&&\times \exp \left\{ i{\text{Str}}[{\mathcal{A}}B]-i\sum_{j}{\text{Str}}%
[\psi _{j}\otimes {\bar{\psi}_{j}}B]-\delta {\text{Str}}[B^{2}]\right\} ,
\notag
\end{eqnarray}%
where $\delta $ is an infinitely small variable that ensures the convergence
of the integral over the variable $B$ in Eq.(\ref{7}); it can be dropped
once the integral over $B$ is convergent. Now, {due to} {the }%
convergence of the integral in Eq. (\ref{7}) and the presence of $\delta $,
we are free to change the order of {the }integration over {the
}supermatrix $B$ and {the }supermatrices $\psi _{i}\otimes {\bar{\psi}%
_{i}}$. {The }integration over {the supervectors }$\psi ,\bar{%
\psi}$ {leads to}

\begin{eqnarray}
&&\int D\psi D{\bar{\psi}}\exp \left\{ -i\sum_{j}{\text{Str}}[\psi
_{j}\otimes {\bar{\psi}}_{j}B]\right\}  \label{8} \\
&=&\int D\psi D{\bar{\psi}}\;e^{i\sum_{j}{\bar{\psi}}_{j}B\psi _{j}}={\text{%
Sdet}}[iB].  \notag
\end{eqnarray}%
Then, the integral over $B$ acquires the form
\begin{equation}
I_{B}=\int_{{{\mathbb{H}}_{n}}}DB\;\;{\text{Sdet}}[iB]\;e^{i{\text{Str}}[{%
\mathcal{A}}B]},  \label{9}
\end{equation}%
where we dropped $\delta $ due to the convergence of the integral Eq. (\ref%
{9}). The integral Eq. (\ref{9}) can be calculated {by} changing the
integration variable $B$ to $B^{\prime }={\mathcal{A}}B$. Supermatrix $%
B^{\prime }$ is not necessarily Hermitian, however it obeys the constraint ${%
\text{Str}}[B^{\prime }]={\text{Str}}\left[ \left( B^{\prime }\right)
^{\dagger }\right] $. By definition $B^{\prime }$ is an element of the
vector space $\Lambda ^{2}({\mathbb{H}}_{n})$ (for definition see Appendox~B). Taking into account the fact that {due to} the supersymmetry the
Berezinian of the transformation $B^{\prime }={\mathcal{A}}B$ is unity and
\begin{equation}
{\text{Sdet}}B=\frac{{\text{Sdet}}B^{\prime }}{{\text{Sdet}}{\mathcal{A}}},
\label{10}
\end{equation}%
we obtain
\begin{eqnarray}
I_{B} &=&{\text{Sdet}}{\mathcal{A}}^{-1}\int_{\Lambda ^{2}({\mathbb{H}}%
_{n})}DB^{\prime }\;{\text{Sdet}}B^{\prime }\;e^{i{\text{Str}}B^{\prime }}
\label{11} \\
&=&C_{n}\;{\text{Sdet}}{\mathcal{A}}^{-1}.  \notag
\end{eqnarray}%
The coefficient $C_{n}$ is calculated in Appendix~B, yielding $C_{n}=1$. As
a result, one arrives at the bosonized representation for the integral Eq. (%
\ref{7})
\begin{equation}
\label{lF}
I_{F}=\int_{{{\mathbb{H}}_{n}}}D{\mathcal{A}}\;\;{\text{Sdet}}{\mathcal{A}}%
^{-1}\;F({\mathcal{A}}).
\end{equation}

To finalize this section we remind {the reader }that in Eq. (\ref{lF}%
) the integration over the linear space of Hermitian supermatrices, $\int_{%
\mathbb{H}_{n}}D{\mathcal{A}}$, with Berezin measure is understood {%
here }as follows: (\emph{i}) First we diagonalize the matrix ${\mathcal{A}}$
and then integrate over the eigenvalues. (\emph{ii}) Integration over
{\textquotedblleft boson-boson\textquotedblright\ eigenvalues} is
performed in the infinite interval $\{-\infty ,\infty \}$, whereas {%
the }integration over {the }\textquotedblleft {%
fermion-fermion\textquotedblright\ }eigenvalues is performed ({in
contrast to Refs.~\onlinecite{YEK,Littelmann}) }in the non-compact interval $%
\{-i\infty ,i\infty \}$. In this way, the integral in Eq.~(\ref{formula})
over supervectors is reduced to an integral over commuting variables.
{It is worth mentioning that the presence of }$\text{Sdet}{\mathcal{A}}^{-1}$%
{\ in Eq. (\ref{formula}}) {leads to a singular product }$%
\prod_{i}\tilde{\lambda}_{i}^{-1}$, {which make the integral very
sensitive to the contour of the integration over the FF-eigenvalues }$\tilde{%
\lambda}_{i}$.

Representing the integral over supervectors in terms of an integral
over the supermatrices is more than just changing the variables of the
integration. Usually, the term bosonization is used for the procedure of a
replacement of an electron model by a model describing collective bosonic
excitations. For example, the traditional $\sigma $-model describes
so called diffusion modes instead of electrons in a random potential. As our
transformation is exact and is based on the supersymmetry, we find it proper
to use the word \textquotedblleft superbosonization\textquotedblright\ for
the transformation, Eq. (\ref{formula}), complemented by the rules of the
integration over the eigenvalues of the supermatrices. 

%%%%%%%%%%%%%%%%%%%%%%%%%%%%%%%%%%%%%%%%%%%%
%%%%%%%%%%%%%%%%%%%%%%%%%%%%%%%%%%%%%%%%%%%%

\section{Superbosonization of random matrices: Correlation functions}

In this section we develop a technique for calculation of various
correlation functions in Random Matrix Theory (RMT), such
%characteristics of random matrices
as the averaged density of states, level-level correlations, eigenfunction
correlations and higher order correlation functions. For this purpose,
without loss of generality, we consider an ensemble of $N$-dimensional
Hermitian matrices, $H=\left\{ H_{ij}\right\} $, with continuously
distributed components. For simplicity let us concentrate on the Gaussian
probability density function,
\begin{equation}
{\mathcal{P}}(H)=\prod_{i,j=1}^{N}P_{ij}(H_{ij}),  \label{Gauss1}
\end{equation}%
with the {distribution }functions $P_{ij}$ %:{\mathbb{C}}\rightarrow
%{\mathbb{R}}^{+}$,
%for
($i,j=1\dots N$) {equal to}
\begin{equation}
P(H_{ij})=\frac{1}{2\pi A_{ij}}\exp \left\{ -\frac{H_{ij}H_{ij}^{\ast }}{%
2A_{ij}}\right\} .  \label{Gauss2}
\end{equation}%
Then Eqs. (\ref{Gauss1}), (\ref{Gauss2}) unambiguously define statistical
properties of the matrix entries as %\begin{eqnarray}
%\label{H}
$\langle H_{ij}\rangle =0$, $\langle H_{ii}^{2}\rangle =A_{0}$, and $\langle
H_{ij}^{2}\rangle =A_{ij}$ for $i\neq j$. %\end{eqnarray}
{The Wigner-Dyson unitary ensemble is obtained putting }$A_{ij}=const$
independent on $i,j.$

\subsection{Correlation functions in the superbosonized representation:
General framework}

We begin with the generating functional for \emph{n-point} correlation
functions
\begin{eqnarray}
Z(J_{1}\ldots J_{n}) &=&\int D\psi D{\psi ^{+}}\;\;\exp \left\{
i\sum_{i,j=1}^{N}{\psi _{i}^{+}}{\mathcal{M}}_{i,j}^{J}\psi _{j}\right\} ,
\notag \\
&&  \label{GF}
\end{eqnarray}%
with the matrix ${\mathcal{M}}_{i,j}^{J}$ is defined as
\begin{eqnarray}
{\mathcal{M}}_{i,j}^{J} &=&L\delta _{ij}{\mathcal{H}}_{j}^{J}+LH_{ij},
\label{GF1} \\
{\mathcal{H}}_{j}^{J} &=&\left( -E+\frac{\omega +i0}{2}\Lambda -J\;\hat{s}%
\right) .  \notag
\end{eqnarray}%
In Eq. (\ref{GF}) $\psi _{i}$ are supervectors with $n$ bosonic and $n$
fermionic components, the source terms , $J_{i}$ ($i=1\ldots n$), in Eq. (%
\ref{GF1}) are real parameters multiplied by diagonal $2n\times 2n$
matrices, $\hat{s}$, which break the fermion-boson (FB) symmetry. Parameter $%
E$ stands for the energy and $\omega $ is the frequency. The $2n$%
-dimensional \emph{supermatrices} $L$, $\Lambda $ and $\hat{s}$ are defined
as
\begin{eqnarray}
L &=&\left(
\begin{array}{cc}
id_{n} & 0 \\
0 & \hat{k}%
\end{array}%
\right) _{FB},\quad \quad \Lambda =\left(
\begin{array}{cc}
\hat{k} & 0 \\
0 & \hat{k}%
\end{array}%
\right) _{FB},  \label{LL} \\
\hat{s} &=&\left(
\begin{array}{cc}
id_{n} & 0 \\
0 & -id_{n}%
\end{array}%
\right) _{FB},\qquad \qquad  \notag
\end{eqnarray}%
with $n$-dimensional unity matrix, $id_{n}$, and $n$-dimensional diagonal
matrix, $\hat{k}=diag(1,-1)$. Purpose of introducing the matrix, $\hat{k}$,
is that it distinguishes between the advanced and the retarded (A/R) Green
functions.

To derive the supersymmetric action for RMT, one has to perform averaging in
the generating functional, $Z(J_{1}\ldots J_{n})$, over realizations of the
entries of the random matrix, $H$. Carrying out such an averaging with the
{probability distribution }defined in Eqs. (\ref{Gauss1}, \ref{Gauss2}%
), one obtains
\begin{eqnarray}
&&\langle Z(J_{1}\ldots J_{n})\rangle =\int D\psi D{\bar{\psi}}\qquad \qquad
\qquad \qquad  \label{1} \\
&&\times \exp \left\{ {i\sum_{i}{\bar{\psi}_{i}}{\mathcal{H}}_{i}^{J}\psi
_{i}-\frac{1}{2}\sum_{i,j}A_{i,j}({\bar{\psi}_{i}}\psi _{j})({\bar{\psi}_{j}}%
\psi _{i})}\right\} ,  \notag
\end{eqnarray}%
where we have defined ${\bar{\psi}_{i}}=\psi _{i}^{+}L$. At this point we
note that for the constituent terms of the action (expressions in exponent),
Eq. (\ref{1}), the following identities hold
\begin{eqnarray}
&&{\bar{\psi}_{i}}{\mathcal{H}}_{i}^{J}\psi _{i}={\text{Str}}\left[ \psi
_{i}\otimes {\bar{\psi}_{i}}\left( -E+\frac{\omega +i0}{2}\Lambda
-J_{i}\right) \right] ,  \notag \\
&&({\bar{\psi}_{j}}\psi _{i})({\bar{\psi}_{i}}\psi _{j})=-{\text{Str}}(\psi
_{i}\otimes {\bar{\psi}_{i}}\psi _{j}\otimes {\bar{\psi}_{j}}),
\label{ident}
\end{eqnarray}

The crucial step towards calculation of {the} correlation functions
in Gaussian random matrix theory under consideration, is the evaluation of
the super-integrals in Eq. (\ref{ident}) from their superbosonized
representation Eq. (\ref{formula}). With the help of {the}
superbosonization formula Eq. (\ref{formula}) we can represent the
generating functional, $\langle Z(J)\rangle $, in the form
\begin{eqnarray}
&&\langle Z(J_{1}\ldots J_{n})\rangle =\int_{{\mathbb{H}_{n}}%
}\prod_{i}DQ_{i}\;{\text{Sdet}}[Q_{i}]^{-1}\;  \label{avZ} \\
&&\times \exp \left\{ i\sum_{i}{\text{Str}}[Q_{i}{\mathcal{H}}_{i}^{J}]-%
\frac{1}{2}\sum_{i,j}A_{i,j}\;{\text{Str}}[Q_{i}Q_{j}]\right\} ,  \notag
\end{eqnarray}%
where each of the integrals over the linear space of complex Hermitian
supermatrices, ${\mathbb{H}_{n}}$, should be performed first diagonalizing
matrices ${Q}_{i}$ and then integrating over their eigenvalues. As was
mentioned in Inroduction, integration over {BB-}eigenvalues is
performed along the real axis, $(-\infty ,\infty )$, whereas integration
over {FF-}eigenvalues is performed along the imaginary axis, $%
(-i\infty ,i\infty )$.

In conclusion of this subsection we note that {the }derivatives of
the averaged generating functional, $\left\langle Z(J_{1}\ldots
J_{n})\right\rangle $, taken at zero source, $J=0$, define the advanced and
retarded Green functions in RMT \cite{VWZ, G}. The $n$-point Green functions
can be expressed via the derivatives of $\left\langle Z(J_{1}\ldots
J_{n})\right\rangle $ functional in a standard way,
\begin{eqnarray}
&&G^{{R/A}}(E_{1}\ldots E_{n}) =\frac{1}{\pi ^{n}}%
\;\left\langle \prod_{i=1}^{n}{\text{Tr}}\left[ \frac{1}{E_{i}-H\pm i0}%
\right] \right\rangle  \notag \\
&=&\frac{1}{(2\pi )^{n}}\;\frac{\partial ^{n}}{\partial J_{1}\cdots \partial
J_{n}}\left\langle Z(J_{1}\ldots J_{n})\right\rangle {\text{{\Huge $|$}}}%
_{J_{i}=0},  \label{GrF}
\end{eqnarray}%
which define the universal characteristics of RMT. {As usual, the
sign }\textquotedblleft $+$\textquotedblright {\ in the denominator
corresponds to the retarded Green function }$G^{R}${, while the sign
\textquotedblleft }$-${\textquotedblright\ to the advanced one }$%
G^{A}.${\ }

\subsection{Correlation functions for diagonal random matrices.}

Let us first show how the method developed here works for diagonal
random matrices. Although this case is not the most interesting one, it
allows one to understand how the method works. We remind the reader that the
conventional non-linear $\sigma $-model \cite{Efetov1,Efetov2,VWZ}
is not applicable in this case. 

For diagonal random matrices we have $A_{ij}=0$ for $i\neq j$, and thus the
averaged generating functional Eq. (\ref{avZ}) acquires the form
\begin{eqnarray}
&&\langle Z_{0}(J_{1}\ldots J_{n})\rangle =\int_{{\mathbb{H}_{n}}%
}\prod_{i}DQ_{i}\;{\text{Sdet}}[Q_{i}]^{-1}\;  \label{0order} \\
&&\times \exp \left\{ i\sum_{i}{\text{Str}}[Q_{i}{\mathcal{H}}_{i}^{J}]-%
\frac{1}{2}\sum_{i}A_{0}{\text{Str}}[Q_{i}^{2}]\right\} ,  \notag
\end{eqnarray}%
where supermatrices ${\mathcal{H}}_{i}^{J}$ are given by Eq. (\ref{GF1}), $%
\hat{s}$ is given by Eq. (\ref{LL}) and $Q_{i}$ are Hermitian supermatrices
with $n$ bosonic and $n$ fermionic entries.

Calculation of $\left\langle Z_{0}(J_{1}\ldots J_{n})\right\rangle $ can be
performed in a similar way, as the calculation of $C_{n}$ in Appendix B.
Namely, first we diagonalize the supermatrices $Q_{i}$, and afterwards
perform IZ-type integration. Since the supermatrices $Q_{i}$ in Eq. (\ref%
{0order}) are Hermitian, they can be diagonalized upon the rotation by the
elements of the unitary supergroup, $SU(n\mid n)$. Substituting
transformation $Q=UQ_{d}U^{+}$, where $U\in SU(n\mid n)$, into Eq. (\ref%
{0order}) we arrive to the following form of the generating functional
\begin{eqnarray}
&&\langle Z_{0}(J_{1}\ldots J_{n})\rangle =\int \prod_{i}DU_{i}\;\int
\prod_{\alpha }D\lambda _{i,\alpha }\;D\tilde{\lambda}_{i,\alpha }\;
\label{0order1} \\
&&\times \Delta ^{2}\left( \{{\tilde{\lambda}_{i,\alpha },}\lambda
_{i,\alpha }\}\right) \prod_{j,\alpha }\;\frac{\lambda _{j,\alpha }\pm i0}{%
\tilde{\lambda}_{j,\alpha }\pm i0}\;  \notag \\
&&\times \exp \left\{ i\sum_{i}{\text{Str}}[Q_{i,d}U_{i}{\mathcal{H}}%
_{i,d}^{J}U_{i}^{+}]-\frac{1}{2}\sum_{i,\alpha }A_{0}(\lambda _{i,\alpha
}^{2}-\tilde{\lambda}_{i,\alpha }^{2})\right\} ,  \notag
\end{eqnarray}%
where the integration over bosonic eigenvalues of $Q_{i}$, $\lambda
_{i,\alpha }$ $(\alpha =1,\cdots n)$, should be carried out along the real
axis, $(-\infty ,\infty )$, and the integration over fermionic eigenvalues, $%
\tilde{\lambda}_{i,\alpha }$ $(\alpha =1,\cdots n)$, should be carried out
along the imaginary axis, $(-i\infty ,i\infty )$. {The
infinitesimally small terms }$\pm i0${\ in Eq. (\ref{0order1}) arise
after removing }$\pm i0${\ from }$H_{i}^{J}${\ in Eq. (\ref%
{0order}) by shifting the variable of the integration }$Q${. }%
Berezinian, $\Delta ^{2}\left( \{{\tilde{\lambda}_{j,\alpha },}\lambda
_{j,\alpha }\}\right) $, is the Jacobian of the diagonalization given by
\begin{eqnarray}
&&\Delta \left( \{{\tilde{\lambda}_{j,\alpha },}\lambda _{j,\alpha
}\}\right) =\prod_{\alpha ,\beta =1}^{n}\frac{(\lambda _{j,\alpha }-\lambda
_{j,\beta })(\tilde{\lambda}_{j,\alpha }-\tilde{\lambda}_{j,\beta })}{%
(\lambda _{j,\alpha }-\tilde{\lambda}_{j,\beta })}  \label{ber} \nonumber\\
&=&\det_{\alpha ,\beta }\left[ \frac{1}{\lambda _{j,\alpha }-\tilde{\lambda}%
_{j,\beta }}\right] .  
\end{eqnarray}

It is transparent that the zero order generating functional Eq. (\ref%
{0order1}) has a factorized form and can be represented as
\begin{equation}
\label{factorization}
\langle Z_{0}(J_{1}\ldots J_{n})\rangle =\left[ \mathcal{Z}_{0}(J)\right]
^{N},
\end{equation}%
where
\begin{eqnarray}
&&\mathcal{Z}_{0}(J)=\int \prod_{i}DU\;\int \prod_{\alpha }D\lambda _{\alpha
}\;D\tilde{\lambda}_{\alpha }  \label{fac} \\
&&\;\times \Delta ^{2}\left( \{{\tilde{\lambda}}_{\alpha },\lambda _{\alpha
}\}\right) \times \prod_{\alpha =1}^{n}\frac{\lambda _{\alpha }\pm i0}{%
\tilde{\lambda}_{\alpha }\pm i0}\;  \notag \\
&&\times \exp \left\{ i{\text{Str}}[Q_{d}U{\mathcal{H}}_{d}^{J}U^{+}]-\frac{1%
}{2}\sum_{\alpha }A_{0}(\lambda _{\alpha }^{2}-\tilde{\lambda}_{\alpha
}^{2})\right\} .  \notag
\end{eqnarray}%
We see that the calculation of the generating functional for diagonal random
matrices reduces to the calculation of the IZ integral. This integral can be
calculated employing the result of Section II for the unitary supergroup,
\cite{G,GW} $U\in SU(n,n)$:
\begin{eqnarray}
I &=&\int DU\;\exp \left\{ i{\text{Str}}[Q_{d}U{\mathcal{H}}%
_{d}^{J}U^{+}]\right\} \qquad \qquad \qquad  \label{IZ1} \\
&=&\left[ 1-\eta \left( \{{\tilde{h}_{\alpha },}h_{\alpha }\}\right) \right]
\;\frac{\prod_{\alpha }\delta (\lambda _{\alpha })\;\delta ({\tilde{\lambda}%
_{\alpha }})}{\Delta ^{2}\bigl(\{{\tilde{\lambda}}_{\alpha },\lambda
_{\alpha }\}\bigr)}  \notag \\
&&+\frac{1}{2^{n(n-1)}\pi ^{n}n!^{2}}\;\frac{\det_{\alpha ,\beta }\left[
e^{ih_{\alpha }\lambda _{\beta }}\right] \;\;\det_{\alpha ,\beta }\left[
e^{ih_{\alpha }\lambda _{\beta }}\right] }{\Delta \bigl(\{{\tilde{h}_{\beta
},}h_{\alpha }\}\bigr)\Delta \bigl(\{{\tilde{\lambda}_{\beta },}\lambda
_{\alpha }\}\bigr)},  \notag
\end{eqnarray}%
where the components $h_{\alpha }$ and ${\tilde{h}}_{\alpha },\;(\alpha
=1\cdots n)$ are {BB} and {FF} eigenvalues of ${\mathcal{H}}%
_{d}^{J},$ respectively, and {the }$\Delta $-functions are defined by
Eq. (\ref{ber}). The boundary term, $\eta $, is given by Eq. (\ref{17}) and
reads
\begin{equation}
\eta \bigl(\{{\tilde{h}_{\beta },}h_{\alpha }\}\bigr)=\frac{\det \left[ \mu %
\bigl({\tilde{h}_{\beta },}h_{\alpha }\bigr)\right] _{\alpha ,\beta =1,\dots
,n}}{2^{n(n-1)}\pi ^{n}\Delta \bigl(\{{\tilde{h}_{\beta },}h_{\alpha },{\}%
\bigr)}},  \label{ueta}
\end{equation}

Here, the matrix $\mu _{\alpha \beta }=\mu \bigl({\tilde{h}_{\beta },}%
h_{\alpha }\bigr)$ is given by
\begin{eqnarray}
&&\mu \bigl({\tilde{h}_{\beta },}h_{\alpha }\bigr) =\int_{-\infty }^{\infty
}D\lambda \int_{-i\infty }^{i\infty }D{\tilde{\lambda}}\frac{1}{(\lambda -{%
\tilde{\lambda}})}\qquad \qquad  \label{ueta1} \\
&&\times \exp \left\{ -\frac{A_{0}}{2}(\lambda ^{2}-{\tilde{\lambda}}%
^{2})+ih_{\alpha }\lambda -i{\tilde{h}}_{\beta }{\tilde{\lambda}}\right\}
\notag \\
&=&\left\{
\begin{array}{ccc}
-\frac{2\pi }{h_{\alpha }-{\tilde{h}}_{\beta }}\;\exp \left\{ -\frac{%
h_{\alpha }^{2}-{\tilde{h}}_{\beta }^{2}}{2A_{0}}\right\} & \text{for} &
h_{\alpha }\neq {\tilde{h}}_{\beta }, \\
&  &  \\
0 & \text{for} & \;\;h_{\alpha }={\tilde{h}}_{\beta }.%
\end{array}%
\right.  \notag \\
&&  \notag
\end{eqnarray}

Now, with the help of the IZ integral, Eq. (\ref{IZ1}), we can perform
integration over $U$ and $U^{+}$,
namely the parameter space of
the unitary supergroup, in the expression
for $\mathcal{Z}_{0}(J)$, Eq. (\ref{fac}). Then, taking into account the
determinant form of the super-Vandermonde determinant Eq. (\ref{ber}),
{we obtain}
\begin{eqnarray}
&&\!\!\!\mathcal{Z}_{0}(J) =\left[ 1-\eta (\{h_{\alpha },{\tilde{h}_{\alpha }}\})%
\right] +\frac{1}{2^{n(n-1)}\pi ^{n}\;\Delta \bigl(\{h_{\alpha },{\tilde{h}%
_{\beta }\}\bigr)}}  \label{ZZ}\nonumber \\
&&\times \det \Biggl[\int_{-\infty }^{\infty }D\lambda \int_{-i\infty
}^{i\infty }D{\tilde{\lambda}}\frac{\lambda }{{\tilde{\lambda}}(\lambda -{%
\tilde{\lambda}})}   \\
&&\times \exp \left\{ -\frac{A_{0}}{2}(\lambda ^{2}-{\tilde{\lambda}}%
^{2})+ih_{\alpha }\lambda -i{\tilde{h}}_{\beta }{\tilde{\lambda}}\right\} %
\Biggr]_{\alpha ,\beta =1\ldots n}.  \notag \\
&&  \notag
\end{eqnarray}%
With this expression for $\mathcal{Z}_{0}(J)$ we are ready to calculate one
and two point (both level-level and eigenfunction-eigenfunction) correlation
functions for Gaussian ensemble of unitary diagonal random matrices. These
calculations are presented in the next two subsections.

\subsubsection{Density of states for diagonal random matrices}

The averaged density of states is expressed in terms of the imaginary part
of the one-point Green function, $G^{A}(E)$, as follows
\begin{equation}
\rho (E)=\frac{1}{\pi }\;{\text{Im}}G^{A}(E).  \label{rho}
\end{equation}

The function $G^{A}(E)$ is related to the averaged generating functional via
Eq. (\ref{GrF}). Employing the factorization property Eq. (\ref%
{factorization}) for one point Green function, one is led to evaluate the
integral in Eq. (\ref{ZZ}) for $n=1$, which means that all the supermatrices
are two dimensional and thus have one bosonic and one fermionic eigenvalue.
Then the Bosonic eigenvalue of the supermatrix ${\mathcal{H}}_{d}^{J}$ will
have the form $h=E+\omega +J$, while the fermionic eigenvalue will have the
form ${\tilde{h}}=E+\omega -J$. Without loose of generality we can set $%
\omega =0$.

For one point Green function one has to take a derivative of the generating
functional, $G_{0}^{A}(E)=(1/2\pi )\partial \left\langle
Z_{0}(J)\right\rangle /\partial J|_{J=0}$, which, as follows from Eqs. (\ref%
{GrF})-(\ref{0order}), can be equivalently represented as $%
G_{0}^{A}(E)=i\langle {\text{Str}}[\hat{s}Q]\rangle $. For $n=1$ the
supersymmetric Vandermonde determinant simplifies and acquires the form $%
\Delta (h,{\tilde{h}})=1/(h-{\tilde{h}})={1}/(2J)$. From here one can easily
realize, that only ${1}/{\Delta ({\tilde{h},}h)}$ term in Eq. (\ref{ZZ})
{contributes} into the derivative in the $J=0$ limit. Therefore, the
expression for one point Green {function}
% function in zero order over $b^2$
{takes the form}
\begin{eqnarray}
\label{DOS2}
G_{0}^{A}(E)\!\! &=&\!\!-\frac{2E}{A_{0}}+\frac{1}{\pi }\int_{-\infty
}^{\infty }D\lambda \int_{-i\infty }^{i\infty }D{\tilde{\lambda}}\;\frac{%
\lambda }{({\tilde{\lambda}}-i0)(\lambda -{\tilde{\lambda}})}\;  \nonumber
\\
&\times &\!\!\exp \left\{ -\frac{A_{0}}{2}(\lambda ^{2}-{\tilde{\lambda}}%
^{2})+iE(\lambda -{\tilde{\lambda}})\right\} . 
\end{eqnarray}%
Evaluation of the integral in Eq. (\ref{DOS2}), presented in Appendix C,
leads to%
\begin{equation}
G_{0}^{A}(E)=\sqrt{\frac{\pi }{2A_{0}}}\left( i+\text{{\large erfi}}\left[
\frac{E}{\sqrt{\;2A_{0}}}\right] \right) e^{-\frac{E^{2}}{2A_{0}}},
\label{DOS3}
\end{equation}%
{where }{\large erfi}$\left( x\right) $ {is the imaginary
error function}%
\begin{equation}
\text{{\large erfi}}\left( x\right) =\frac{2}{\sqrt{\pi }}\sum_{n=0}^{\infty
}\frac{x^{2n+1}}{\left( 2n+1\right) n!}.  \label{e7}
\end{equation}%
{\ }Eq. (\ref{DOS3}) exactly reproduces the averaged advanced Green
function of the Gaussian unitary ensemble of diagonal random matrices (see
for example Refs. \onlinecite{OK1, OK2}). Substituting Eq. (\ref{DOS3}) into
Eq. (\ref{rho}) we find the density of states $\rho _{0}(E)$ for diagonal
random matrices,
\begin{equation}
\rho _{0}(E)=\frac{N^{2}}{\sqrt{2\pi A_{0}}}\;e^{-\frac{E^{2}}{2A_{0}}}.
\label{rho2}
\end{equation}

\subsubsection{Two point correlation function for diagonal random matrices}

In this subsection we show how the superbosonization formula with the flat
integration measure, as defined above, works for four-dimensional
supermatrices. Namely, we employ the developed technique of superbosonized
generating functional Eqs. (\ref{avZ}), (\ref{GrF}) for calculation of
{a two-level} correlation function of diagonal random matrices. For
simplicity we will concentrate on a level-level correlation function having
the following form
\begin{eqnarray}
&&K_{0}^{A}(E_{1},E_{2}) =\frac{1}{\pi ^{2}}\left\langle \text{Tr}\left[
\frac{1}{E_{1}-H-i0}\right]\right.   \label{2pGF} \\
&&\left.\times \text{Tr}\left[ \frac{1}{E_{2}-H-i0}\right]\right\rangle
\!-N(N-1)G_{0}^{A}(E_{1})G_{0}^{A}(E_{2})  \notag
\end{eqnarray}%
where $\mathbf{N}$ is the size of the matrices. We are
aware of the fact that the correlation function $K_{0}^{A}(E_{1},E_{2})$
containing the product of two advanced Green functions is not the
most interesting function characterizing the level correlations. However,
the computation of this function presented here serves merely as a
demonstration of how the method works. We emphasize that the method of
integration adopted in Refs.~\onlinecite{YEK, Littelmann} does not work when
applied to this problem. 

For the case of diagonal random matrices the two-point function Eq. (\ref%
{2pGF}) can be derived upon evaluating Eqs. (\ref{avZ}), (\ref{GrF}). This
can be done making use of the factorization property Eq. (\ref{factorization}%
) with $\mathcal{Z}_{0}(J)$ given by (\ref{ZZ}). The calculation is
straightforward. Since the Vandermonde determinant, $\Delta \bigl(\{{\tilde{h%
}_{\beta },}h_{\alpha }{\}\bigr)}$, in Eqs. (\ref{ueta}) and (\ref{ZZ}) is
always inverse proportional to the source terms, $J_{1}$ and $J_{2}$, it is
easy to see, that only the $\left[ \Delta \bigl(\{h_{\alpha },{\tilde{h}%
_{\beta }\}\bigr)}\right] ^{-1}$ term will contribute to double derivative
in Eq. (\ref{GrF}) taken at $J_{1}=J_{2}=0$. The double derivative of the
Vandermonde determinant is equal {to}
\begin{equation}
{\partial _{J_{1}}}{\partial _{J_{2}}}\left[ \Delta \bigl(\{{\tilde{h}%
_{\beta },}h_{\alpha },{\}\bigr)}\right] ^{-1}\Big\vert_{J_{1},J_{2}=0}=4
\label{e4}
\end{equation}
Therefore, we have {for the function }$K_{0}^{A}(E_{1},E_{2})${%
\ }
\begin{eqnarray}
&&\frac{1}{N}K_{0}^{A}(E_{1},E_{2}) =-\frac{1}{\pi ^{2}}\det \left[ \mu %
\bigl(\{h_{\alpha },{\tilde{h}_{\beta }}\}\bigr)\right] _{\alpha ,\beta =1,2}
\label{2pGF2} \\
&&+\frac{1}{\pi ^{2}}\det \Biggl[\int_{-\infty }^{\infty }D\lambda
\int_{-i\infty }^{i\infty }D{\tilde{\lambda}}\frac{\lambda }{({\tilde{\lambda%
}}+i0)(\lambda -{\tilde{\lambda}})}  \notag \\
&&\times \exp \left\{ -\frac{A_{0}}{2}(\lambda ^{2}-{\tilde{\lambda}}%
^{2})+ih_{\alpha }\lambda -i{\tilde{h}}_{\beta }{\tilde{\lambda}}\right\} %
\Biggr]_{\alpha ,\beta =1,2},  \notag
\end{eqnarray}%
where $\mu \bigl(\{{\tilde{h}_{\beta },}h_{\alpha }\}\bigr)$ is defined by
Eq. (\ref{ueta1}) and $h_{1,2}={\tilde{h}}_{1,2}=E_{1,2}$.

Analysis of the integral over $\lambda $ and $\tilde{\lambda}$ under {%
the }second determinant in Eq. (\ref{2pGF2}) is presented in Appendix C.
%By using a decomposition $\frac{\lambda} {({\tilde\lambda}+i
%0)(\lambda-{\tilde \lambda})} = \frac{1}{{\tilde \lambda} +i 0}
%+\frac{1}{\lambda-{\tilde \lambda}}$ one can represent it as a
Result of the integration can be represented as the sum, $\tilde{\eta}\bigl(%
\{h_{\alpha },{\tilde{h}_{\beta }}\}\bigr)+G_{0}(\{h_{\alpha },{\tilde{h}%
_{\beta }}\})$, where
\begin{equation}
G_{0}(\{{\tilde{h}_{\beta },}h_{\alpha }\})=\sqrt{\frac{\pi }{2A_{0}}}\left(
i+\text{{\large erfi}}\left[ \frac{\tilde{h}_{\beta }}{\sqrt{\;2A_{0}}}%
\right] \right) e^{-\frac{h_{\alpha }^{2}}{2A_{0}}}.  \label{fin}
\end{equation}%
Substituting now Eqs. (\ref{ueta}) and (\ref{fin}) into the determinants in
Eq. (\ref{2pGF2}), {we come to the result}
\begin{equation}
K_{0}^{A}(E_{1},E_{2})=\frac{N}{\pi }\;\left[ \frac{%
G_{0}^{A}(E_{2})-G_{0}^{A}(E_{1})}{E_{1}-E_{2}}\right] ,  \label{e5}
\end{equation}%
where the function $G_{0}^{A}(E)$ is given by Eq. (\ref{DOS3}). 
The result for the level-level correlation function for diagonal random matrices, Eq.~(\ref{e5}), coinciding with Eq.~(\ref{2pGF}), together with
 Eq. (\ref{DOS3}) agrees with the one found in Ref.~\onlinecite{Bovier}. 

\subsection{Non-diagonal contributions to the density of states for almost
diagonal matrices}

In order to show how {the} superbosonization {technique }works
for {less trivial }random matrix theories, we calculate{\ in
this section a }correction to the density of states in the model of almost
diagonal matrices\cite{OK1,OK2} up to the second order {in} {%
the }bandwidth, $b$. By definition, statistical properties of {non-}%
diagonal matrices are described by a single, always positive function, $%
\mathcal{F}(r)$, as $A_{ij}=b^{2}\mathcal{F}\left( |i-j|\right) ,\;\;i\neq j$%
. Function $\mathcal{F}(r)$ can adopt any form provided that it has a
maximum at the center of the band, $r=0$, and decays with the bandwidth, $b$%
, as $r$ becomes large. For small $b$ we have {the }ensemble of
{almost} diagonal random matrices, while for large $b$ we approach
{the} {Wigner-Dyson }Gaussian {Unitary Ensemble (GUE).}

We consider the case of %$A_{ij}=b^2 {\cal F}(|i-j|)$, where
$b\ll 1$ {when the standard non-linear }$\sigma $-{model is
not applicable}. Then, expanding the exponent in Eq. (\ref{avZ}) {in}
$b$, {we have}
\begin{equation}
\langle Z(J)\rangle =\langle Z_{0}(J)\rangle +b^{2}\langle Z_{1}(J)\rangle ,
\label{e6}
\end{equation}%
where zero order {in} $b^{2}$ contribution, $\langle Z_{0}(J)\rangle $%
, corresponds to diagonal random matrices considered in {the }%
previous subsection. Technically, calculation of the correction, $%
b^{2}\langle Z_{1}(J)\rangle $, is similar to {that }of $\langle
Z_{0}(J)\rangle $. It is determined by the form of $A_{ij}$ for {%
almost} diagonal matrices as follows
\begin{eqnarray}
&&\langle Z_{1}(J)\rangle =\frac{1}{2}\sum_{i,j}\mathcal{F}\left(
|i-j|\right) \left\langle {\text{Str}}[Q_{i}Q_{j}]\right\rangle =
\label{1order} \\
&&-\frac{1}{2}\int_{{\mathbb{H}_{n}}}\prod_{i}DQ_{i}\;{\text{Sdet}}%
[Q_{i}]^{-1}\frac{1}{2}\sum_{i,j}\mathcal{F}\left( |i-j|\right) \;{\text{Str}%
}[Q_{i}Q_{j}]\;  \notag \\
&&\times \exp \left\{ i\sum_{i}{\text{Str}}[Q_{i}{\mathcal{H}}_{i}^{J}]-%
\frac{1}{2}\sum_{i}A_{0}\;{\text{Str}}[Q_{i}^{2}]\right\} ,  \notag
\end{eqnarray}%
where, as usual, integration goes over the linear space ${{\mathbb{H}_{n}}}$
with the flat measure. Then, {the} correction to the advanced Green
function, $b^{2}G_{1}^{A}(E)$, is expressed in terms of the correction, $%
b^{2}\langle Z_{1}(J)\rangle $, to the averaged generating functional,
\begin{eqnarray}
&&G_{1}^{A}(E)=\frac{\partial \langle Z_{1}(J)\rangle }{\partial J}
\label{1oder1} \\
&=&\frac{i}{2}\sum_{k,i\neq j}\mathcal{F}\left( |i-j|\right) \;\left\langle {%
\text{Str}}[{\hat{s}}Q_{k}]\;{\text{Str}}[Q_{i}Q_{j}]\right\rangle .  \notag
\end{eqnarray}%
In Eq. (\ref{1oder1}) the averaging, $\langle \ldots \rangle $, is defined
{as}
\begin{eqnarray}
&&\langle F[Q]\rangle =\int DQ\;{\text{Sdet}}[Q]^{-1}\;F[Q]\;\qquad \qquad
\label{notion} \\
&&\times \exp \left\{ i\sum_{i}{\text{Str}}[Q{\mathcal{H}}^{J}]-\frac{1}{2}%
A_{0}\;{\text{Str}}[Q^{2}]\right\} .  \notag
\end{eqnarray}%
Averaging {in} the right hand side of Eq. (\ref{1oder1}) can be
performed using the identity, $\langle Q\rangle =\left\langle \frac{1}{2}{%
\text{Str}}[{\hat{s}}Q]\right\rangle {id}_{2n}$. Next, we make use of this
identity to represent the average in Eq. (\ref{1oder1}) as
\begin{eqnarray}
&&\langle {\text{Str}}[\hat{s}Q_{k}]\rangle _{Q_{k}}\;\langle {\text{Str}}%
[Q_{i}Q_{j}]\rangle _{Q_{i,j}}  \label{prove} \\
&=&\langle {\text{Str}}[{\hat{s}}Q]\rangle _{Q}\;\left\langle \frac{1}{2}{%
\text{Str}}[{\hat{s}}Q]{\text{Str}}[Q]\right\rangle _{Q}\delta _{ki}\;\;\;(%
\text{or}\;\delta _{kj})  \notag \\
&=&-\frac{i}{2}\;G_{0}(E)\;\frac{\partial \langle {\text{Str}}[Q]\rangle ^{J}%
}{\partial J}\delta _{ki}\;\;\;(\text{or, alternatively }\;\delta _{kj}),
\notag
\end{eqnarray}%
where, according to Eq. (\ref{notion}), we have
\begin{eqnarray}
&&\frac{\partial \langle {\text{Str}}[Q]\rangle ^{J}}{\partial J} =\frac{%
\partial }{\partial J}\int DQ\;{\text{Sdet}}[Q]^{-1}\;{\text{Str}}[Q]\;
\label{1order2} \\
&&\times \exp \left\{ i\sum_{i}{\text{Str}}[QH^{J}]-\frac{1}{2}A_{0}\;{\text{%
Str}}[Q^{2}]\right\} .  \notag
\end{eqnarray}

As described above, now again, one has to diagonalize the supermatrix $Q$
and reduce the expression Eq. (\ref{1order}) to IZ integral. For that
purpose, we first notice that the only difference between the expressions
for $\langle {\text{Str}}[Q]\rangle ^{J}$ and $Z_{0}(J)$ is the presence of
the term ${\text{Str}}[Q]$ under integral, which, after diagonalization for
the one-point Green functions ($n=1$ case), produces an additional $\lambda -%
{\tilde{\lambda}}$ term under the integral in Eq. (\ref{ZZ}). Secondly, the
boundary $1-\eta $ term does not contribute here, because the presence of $%
\delta $ functions in Eq. (\ref{IZ1}) together with $\lambda -{\tilde{\lambda%
}}$ in the integral makes it zero. Repeating now the calculation for $%
G_{0}(E)$ {and} keeping in mind the two observations above, one
{finds}
\begin{eqnarray}
\frac{\partial \langle {\text{Str}}[Q]\rangle ^{J}}{\partial J} &=&\frac{1}{%
\pi }\int\limits_{-\infty }^{\infty }D\lambda \int\limits_{-i\infty
}^{i\infty }D{\tilde{\lambda}}\;\frac{\lambda }{({\tilde{\lambda}}+i0)}\;
\label{1order3} \\
&&\times \exp \left\{ -\frac{A_{0}}{2}(\lambda ^{2}-{\tilde{\lambda}}%
^{2})+iE(\lambda -{\tilde{\lambda}})\right\}  \notag \\
&=&\frac{2i}{A_{0}}\;N\;\mathcal{R}_{N}[\mathcal{F}]\;G_{0}^{A}(E).  \notag
\end{eqnarray}%
Substituting now Eqs. (\ref{prove}) and (\ref{1order3}) into the expression
for the first order correction to the Green function $G_{1}(E)$, Eq. (\ref%
{1oder1}), we obtain
\begin{equation*}
G_{1}(E)=N\;\mathcal{R}_{N}[\mathcal{F}]\;G_{0}^{A}(E)\;\frac{1}{A_{0}}\;%
\left[ G_{0}^{A}(E)-1\right] ,
\end{equation*}%
where $\mathcal{R}_{N}[\mathcal{F}]\equiv 2\sum_{l=1}^{N}\mathcal{F}(l)$.
Then, for the first order {in} $b^{2}$ correction to density of
states, $\rho _{1}(E)=\pi ^{-1}\text{Im}G_{1}^{A}(E)$, one easily finds
\begin{eqnarray}
\rho _{1}(E) &=&\mathcal{R}_{N}[\mathcal{F}]\;\rho _{0}(E)\;\frac{1}{A_{0}}%
\;\left( E\;\sqrt{\frac{2\pi }{A_{0}}}\;\text{{\large erfi}}\left[ \frac{E}{%
\sqrt{2A_{0}}}\right] -1\right) ,  \label{ro} \nonumber\\
&&  
\end{eqnarray}%
which exactly reproduces the results first obtained with the help of the
virial expansion\cite{OK1,OK2}.

\section{Outlook}

We have presented a new scheme of computations using the
superbosonization formula, Eq. (\ref{formula}), first proposed in Ref. \onlinecite%
{E}. We have proven that this formula is exact and have given a precise
recepy for the performing integration for many point correlation functions
for the unitary ensemble. In contrast to a previous study \cite%
{YEK,Littelmann} the integration over the eigenvalues in the fermion-fermion
block of the supermatrices is performed from $-i\infty $ {to }$%
i\infty $ and not along a circle. This way of the
integration has allowed us to obtain regular integrals and calculate them in
several cases. 

The proof of our approach and proposed method of computation of the
integrals is heavily based on the supersymmetric extension of the
Itzykson-Zuber integral. This integral in known only for systems with broken
time-reversal symmetry (unitary ensemble) and this why we consider here only
such systems. At the same time, the proposed method of the integration over
the eigenvalues of the supermatrices when one integrates over the
eigenvalues in the boson-boson block from $-\infty $ {to} $\infty $
and over the eigenvalues in the fermion-fermion block from $%
-i\infty $ {to }$i\infty $ looks very general. This
encourages us to make a guess that this way of the integration can also be
used for time reversal invariant ensembles. Of course, such a guess must be
checked and proven in the future. 

We have demonstrated that the application of the bosonization formula to
random band matrix (RBM)\cite{Fyodorov1,MPhysRep,FM,FM1,Cuevas,YO}
 models with small bandwidth $b$ reproduces
the perturbative expansion of DOS obtained by virial expansion \cite{OK1}.
We have also computed the simplest two-point correlation function
containing a product of two advanced Green functions for the
ensemble of diagonal matrices. Of course, calculating an average
product of both retarded and advanced Green functions would be a more
interesting task but we leave it for future study. It is important at the
moment that our method allows us to calculate many-point correlations
functions for cases where the way of the integration developed in Refs. \cite%
{YEK,Littelmann} does not work. We have made comparison with the known
results only for checking our approach and demonstration of details of the
computation.

Eq. (\ref{formula}) complemented by our recipe of the integration is
exact and most general representation of the integrals over supervectors in
terms of integrals over supermatrices. The traditional non-linear $\sigma $-{model, Eqs. (\ref{e0})
 and (\ref{e1}), can be obtained using the saddle-point approximation for
calculation of the integral over the supermatrix }$Q$ {is less
general. Taking into account a success of the latter in solving numerous
problems (see, e.g. Ref.~\onlinecite{Efetov1}) we believe that its generalization
can also bring new interesting results. }

Finally, we would like to mention that to this point 
disordered systems have been actually successfully studied using  
supersymmetric $\sigma $-model (including statistical properties of the energy levels in small metallic disordered grains),
and we mostly focused here on a field theory for random matrix ensembles and non-perturbative effects therein.
Another field of great interest of course is the non-perturbative study of various correlation functions in strongly interacting systems. Examples of such systems that potentially can be studied non-perturbatively using superbosonization include among others {\em (i)} the field theory of many-body localization in random spin chains\cite{Altland}; {\em (ii)} quantum phase transitions at the boundary of topological superconductors in two and three dimensions, which have been argued to support supersymmetry at long distances and times\cite{grover}.

\begin{acknowledgments}
T.A.S. acknowledges startup funds from UMass Amherst.
K.B.E. gratefully acknowledges the financial support of the Ministry of
Education and Science of the Russian Federation in the framework of Increase
Competitiveness Program of NUST~\textquotedblleft MISiS\textquotedblright\
(Nr.~K2-2014-015).
\end{acknowledgments}

\appendix

\begin{widetext}

\section{Integral over ${\mathbb{H}}_n$}

% {\em Identity 1}:
Let ${\mathbb{H}}_n$ is the linear space of Hermitian
 $2n\times 2n$ supermatrices.
Then for all ${\mathcal A}\in{\mathbb{H}}_n$ the convergent
integral,
\begin{eqnarray}
\label{6}  {\hat \Upsilon}({\mathcal A})=\lim_{\eta \rightarrow 0}
\int_{{\mathbb{H}}_n} DB\exp\left\{i {\text{Str}} [{\mathcal A}B]-
\tilde{\eta} {\text{Str}}[B^2]\right\},
\end{eqnarray}
taken over ${\mathbb{H}}_n$ with Berezin measure
satisfies the condition
%{\em unambiguously} defines a function
\begin{eqnarray}
\label{delta} \int_{{\mathbb{H}}_n} D{\mathcal A}^{\prime}\; {\hat
\Upsilon} ({\mathcal A}^{\prime}-{\mathcal A})\equiv 1.
\end{eqnarray}
%Here, as in Eq. (\ref{6}), the integral is taken over the linear
%space of Hermitian $2n\times 2n$ supermatrices with flat measure.
Moreover, for any map, ${\mathcal
F}:{\mathbb{H}}_n\rightarrow{\mathbb{G}}$, that converges
exponentially (or faster), the identity
\begin{eqnarray}
\label{id2'} {\mathcal F}(Q)\equiv
\int_{{\mathbb{H}}_n}D{\mathcal A}{\mathcal F}({\mathcal
A}){\hat \Upsilon} ({\mathcal A}-Q)
\end{eqnarray}
always holds.

To derive identities (\ref{delta}) and (\ref{id2'}) for ${\hat
\Upsilon} ({\mathcal A})$,  one can first formally perform
integration in the definition Eq. (\ref{6}) of ${\hat \Upsilon}
({\mathcal A})$,
 under the limit. This integration yields
\begin{eqnarray}
 \label{A-1}
{\hat \Upsilon} ({\mathcal A})=\lim_{\eta \rightarrow 0}
\Big[\frac{1}{4 \pi \eta} \Big]^{\frac{N}{2}}\exp\left\{
-\frac{Str {\mathcal A}^2}{4 \eta}\right\},
\end{eqnarray}
where the limit is well defined. With the help of Eq. (\ref{A-1}),
the right hand side of Eq. (\ref{id2'}) can be represented as
follows
\begin{eqnarray}
\label{A-2} &&\int_{{\mathbb{H}}_n}D{\mathcal A}{\mathcal
F}({\mathcal A}){\hat \Upsilon} ({\mathcal
A}-Q)=\int_{{\mathbb{H}}_n}D{\mathcal A}{\mathcal F}({Q+\mathcal
A}){\hat \Upsilon} ({\mathcal A})\nonumber \\
&=&\lim_{\eta \rightarrow 0}\Big[\frac{1}{4 \pi \eta}
\Big]^{\frac{N}{2}} \int_{{\mathbb{H}}_n}D{\mathcal
A}\Big\{{\mathcal F}(Q)+ tr[{\mathcal F}^{\prime}({\mathcal
A}){\mathcal A}]+\cdots +tr[{\mathcal F}^{(n)}({\mathcal
A}){\mathcal A}^n]+\cdots \Big\}
\exp\Big\{ -\frac{Str {\mathcal A}^2}{4 \eta}\Big\}=\nonumber \\
&=&{\mathcal F}(Q)+\lim_{\eta \rightarrow 0}\Big[\frac{1}{4 \pi
\eta} \Big]^{\frac{N}{2}} \int_{{\mathbb{H}}_n}D{\mathcal
A}\Big\{tr[{\mathcal F}^{\prime}({\mathcal A}){\mathcal A}]+\cdots
+tr[{\mathcal F}^{(n)}({\mathcal A}){\mathcal A}^n]+\cdots \Big\}
\exp\Big\{ -\frac{Str {\mathcal A}^2}{4 \eta}\Big\},
\end{eqnarray}
where we have Tailor expanded the function ${\mathcal
F}({Q+\mathcal A})$ around $\mathcal A=0$. We note, that such
an expansion exists due to the specific constraints on the
function $\mathcal F$, outlined in Section III.
 To finalize our proof, it is left to show that
\begin{eqnarray}
 \label{A-3}
\lim_{\eta \rightarrow 0}\Big[\frac{1}{4 \pi \eta}
\Big]^{\frac{N}{2}} \int_{{\mathbb{H}}_n}D{\mathcal A}tr[{\mathcal
A}^n] \exp\Big\{ -\frac{Str {\mathcal A}^2}{4 \eta}\Big\}=0.
\end{eqnarray}
Eq. (\ref{A-3}) is proven by introducing the generating
functional,
\begin{eqnarray}
\label{A-4} W_{\eta}(K)=\Big[\frac{1}{4 \pi \eta}
\Big]^{\frac{N}{2}} \int_{{\mathbb{H}}_n}D{\mathcal A}\exp\Big\{
-\frac{Str {\mathcal A}^2}{4 \eta} + Str[K {\mathcal A}]\Big\}
=\exp\{\eta Str[K^2] \},
\end{eqnarray}
and observing that
\begin{eqnarray}
\label{A-5} \lim_{\eta \rightarrow 0}\Big[\frac{1}{4 \pi \eta}
\Big]^{\frac{N}{2}} \int_{{\mathbb{H}}_n}D{\mathcal A}
Str[{\mathcal A}^n] \exp\Big\{ -\frac{Str {\mathcal A}^2}{4
\eta}\Big\}= \lim_{\eta \rightarrow 0}\Big[\frac{1}{4 \pi \eta}
\Big]^{\frac{N}{2}} Str\Big[ \frac{\delta^{n}  W_{\eta}(K)}{\delta
{\mathcal A}_1 \cdots \delta {\mathcal A}_n}\Big] \Big\vert_{K=0}
=0.
\end{eqnarray}

\section{Calculation of $C_n$}

%{\em Identity 2}:
Let the formal sums of Hermitian super-bivectors
(product of two supermatrices, each of them being from
${\mathbb{H}}_n$) constitute a vector space
$\Lambda^2({\mathbb{H}}_n)$ called the second exterior power of
${\mathbb{H}}_n$. Then the integral
\begin{eqnarray}
\label{Cn} C_n=\int_{\Lambda^2({\mathbb{H}}_n)} D
B'\;{\text{Sdet}}B'\;e^{i{\text{Str}}B'}
\end{eqnarray}
over the vector space $\Lambda^2({\mathbb{H}}_n)$ is unity,
$C_n=1$.

According to the proposed prescription, one evaluates integral Eq.~(\ref{Cn}) first by diagonalizing the supermatrix $B^{\prime}$. As
already stated, any given complex $2n \times 2n$ supermatrix
$B^{\prime}$ can be diagonalized by the transformation
$B^{\prime}=UB_d^{\prime}{\bar V}$, where $U\in U(n\mid n)$,
${\bar V}\in {U(n\mid n)}/{U^{2n}(1)}$. Substituting this
transformation into Eq. (\ref{9}) we obtain an integral over the
eigenvalues, $B_d^{\prime}$, and diagonalization "angles" $U$ and
$\bar{V}$; the latter integral is nothing but supersymmetric
Itzikson-Zuber \cite{IZ} integral.

%Below we will present the calculation of supersymmetric IZ
%integral following the article \cite{GW}.

%It is necessary to make following remark here. The "naive"  answer
%$\Gamma_0 \left[\{b_j,{\tilde b_j}\}\mid \{ \lambda _j,{\tilde
%\lambda_j}\}\right]$ for IZ integral was obtained in \cite{G, AM}
%by solving supersymmetric heat equation following the technique
%developed in the original article \cite{IZ}. It is straightforward
%to  check that the boundary term $\propto (1-\eta)$ in (\ref{15})
%also satisfies the heat equation.

In order to evaluate integral in Eq. (\ref{11}), we consider the
following generalized integral
\begin{eqnarray}
\label{18} \int DB\; {\text{Sdet}}[B]\;\exp\left\{\frac{i}{t}\;
{\text{Str}}[Q_dB]\right\}=\int DB\;{\text{Sdet}}[B]
\exp\left\{\frac{1}{2t}\;{\text{Str}}[B^2-Q_d^2]-\frac{1}{2t}\;
{\text{Str}}[(B-iQ_d)^2]\right\} \nonumber\\= \int \prod _i
db_i\;d{\tilde b_i}\; \Delta ^2\left(\{b_j^2,{\tilde
b_j}^2\}\right) \left [\prod_i \frac{{\tilde b_i}}{b_i}\;
\exp\left\{\frac{1}{2t}\; \left[(b_i^2-{\tilde b_i}^2)-(\lambda
_i^2-{\tilde \lambda _i}^2)\right]\right\}\right ]\; \Gamma
\left[\{b_j,{\tilde b_j}\}\mid \{ \lambda _j,{\tilde \lambda
_j}\}\right],\qquad
\end{eqnarray}
which coincides with Eq. (\ref{18}) in the case when $Q_d$ is an
identity matrix. Before setting $Q_d=id$, first let us note
that for any complex $2n\times 2n$ supermatrix of the following
diagonal form:
\begin{eqnarray}
\label{19} \Lambda = \left (
  \begin{array}{cc}
  x \otimes id_n &  0 \\
  0    &  y\otimes id_n \\
  \end{array}
\right ),\qquad x,y \in \mathbb{R},
\end{eqnarray}
where $id_n$ is the $n\times n$ identity matrix, the $\eta$ term
has the form
\begin{eqnarray}
\label{20} \eta _{\Lambda}(x,y) = \left (
1-e^{-{\frac{x^2-y^2}{2t}}}\right )^n.
\end{eqnarray}
The term, $\Gamma _0\left[\{b_j,{\tilde b_j}\}\mid x,y\right],$
[see Eq. (\ref{15})] corresponding to the matrix $\Lambda$
vanishes. This is because the Vandermonde determinant, $\Delta
_{\Lambda}(x,y)$, in the denominator will cancel one of
determinants involving Bessel function in the nominator. However
the next determinant, which is equal to zero, remains. Thus, we
see that if our $2n$ dimensional complex supermatrix $Q_d=id$
(which means $x=y$ above), then the corresponding $\eta$ term
$\eta _{id}(1,1) = \left ( 1-e^{-{\frac{1-1}{2t}}}\right )^n =0$.
Therefore, from Eq.(\ref{15}),
 we obtain
\begin{eqnarray}
\label{id} \Gamma \left[\{b_j,{\tilde b_j}\}\mid \{1\dots
1\}\right]= \frac{\prod _i\delta (b_i)\delta ({\tilde
b_i})}{\Delta ^2\bigl(\{b_j^2,{\tilde b_j}^2 \}\bigr)}.
\end{eqnarray}
Substituting Eq. (\ref{id}) into Eq. (\ref{18}), where as $Q_d$ a
unity matrix is taken with $t=1$, one obtains
\begin{eqnarray}
\label{Cn1} C_n=\int D
B^{\prime}\;{\text{Sdet}}[B^{\prime}]\;e^{i{\text{Str}}B^{\prime}}&=&\int
\prod _i db_i\;d{\tilde b_i}\; \Delta ^2\left(\{b_j^2,{\tilde
b_j}^2\}\right) \left \{\prod_i \frac{{\tilde b_i}}{b_i}\;
e^{\frac{1}{2} \left[b_i^2-{\tilde b_i}^2\right]}\right \}\;
\Gamma \left[\{b_j,{\tilde b_j\} |  \{1\dots 1}\}\right] \nonumber \\
&=&\int \prod _i db_i\;d{\tilde b_i}\;\left \{\prod_i
\frac{{\tilde b_i}}{b_i}\; e^{\frac{1}{2} \left[b_i^2-{\tilde
b_i}^2\right]}\right \}{\prod _i\delta (b_i)\delta ({\tilde
b_i})}=1.
\end{eqnarray}
The last equality holds, since our integration contours are
shifted by an infinitesimal $\delta$ and $i\delta$ with respect to
the imaginary and real axis correspondingly. This completes the
computation of $C_n$.

%%%%%%%%%%%%%%%%%%%%%%%%%%%%%%%%%%%%%%%%

\section{Evaluation of double-integrals}
Here we will evaluate the following integral
\begin{eqnarray}
\label{doubleint}
 {\mathcal I}(h,\tilde{h})= \int_{-\infty}^{\infty}d\lambda\int_{-i\infty}^{i\infty}d{\tilde \lambda}\;\frac{\lambda}
{({\tilde\lambda}-i0)(\lambda-{\tilde
\lambda})}\;\exp\left\{-\frac{A_0}{2}(\lambda^2-{\tilde
\lambda}^2)+ih\lambda-i\tilde{h}\tilde{\lambda }\right\}.
%+i E(\lambda-{\tilde \lambda})\right\}.
\end{eqnarray}
Making use of the decoupling
\begin{eqnarray}
\label{decoup} \frac{\lambda}
{({\tilde\lambda}-i0)(\lambda-{\tilde
\lambda})}=\left(\frac{1}{\tilde\lambda-i0}+\frac{1}{\lambda-\tilde\lambda}\right),
\end{eqnarray}
we represent the double-integral, ${\mathcal I}(h,\tilde{h})$, as
the sum
\begin{eqnarray}
\label{sum} {\mathcal I}(h,\tilde{h})={\mathcal
I}_1(h,\tilde{h})+{\mathcal I}_2(h,\tilde{h}),
\end{eqnarray}
where
\begin{eqnarray}
\label{I1}
 {\mathcal I}_1(h,\tilde{h})= \int_{-\infty}^{\infty}d\lambda\int_{-i\infty}^{i\infty}d{\tilde \lambda}\;\frac{1}
{({\tilde\lambda}-i0)}\;\exp\left\{-\frac{A_0}{2}(\lambda^2-{\tilde
\lambda}^2)+ih\lambda-i\tilde{h}\tilde{\lambda }\right\}
%+i E(\lambda-{\tilde \lambda})\right\}
,\nonumber\\
 {\mathcal I}_2(h,\tilde{h})= \int_{-\infty}^{\infty}d\lambda\int_{-i\infty}^{i\infty}d{\tilde \lambda}\;\frac{1}
{(\lambda-{\tilde
\lambda})}\;\exp\left\{-\frac{A_0}{2}(\lambda^2-{\tilde
\lambda}^2)+ih\lambda-i\tilde{h}\tilde{\lambda }\right\}.
%+i E(\lambda-{\tilde \lambda})\right\}.
\end{eqnarray}
In the following two subsections we will evaluate integrals
${\mathcal I}_1(h,\tilde{h})$ and ${\mathcal I}_2(h,\tilde{h})$
respectively.
\subsection{Calculation of ${\mathcal I}_1$}

In order to evaluate ${\mathcal I}_1(h,\tilde{h})$ we recall that
\begin{eqnarray}
\label{decoup2} \frac{1}
{({\tilde\lambda}-i0)}=\text{\large{P}}\;\frac{1}{\tilde{\lambda}}+i\pi
\delta(\tilde{\lambda})
\end{eqnarray}
where symbol $\text{\large P}$ denotes the principal value of the
integral. Then for ${\mathcal I}_1(h,\tilde{h})$ we have
\begin{eqnarray}
\label{Int1} {\mathcal I}_1(A_0,h,\tilde{h})=i\pi
\int_{-\infty}^{\infty}\exp\left\{-\frac{A_0}{2}\;\lambda^2+ih\lambda\right\}+\tilde{\mathcal
I}_1(A_0,h,\tilde{h}),
\end{eqnarray}
with
\begin{eqnarray}
\label{Itilde1} \tilde{\mathcal I}_1(A_0,h,\tilde{h})=
\int_{-\infty}^{\infty}d\lambda\;
\text{\large{P}}\!\int_{-i\infty}^{i\infty}\frac{d{\tilde
\lambda}}{{\tilde \lambda}}\;
\;\exp\left\{-\frac{A_0}{2}\;(\lambda^2-{\tilde \lambda}^2)
+ih\lambda-i\tilde{h}\tilde{\lambda }\right\}.
%iE(\lambda-{\tilde \lambda})\right\}
\end{eqnarray}
The presence of the principle value in Eq. (\ref{Itilde1}) insures
the possibility of bringing the integral to the Gaussian form
first by taking the derivative over $\tilde{h}$:
\begin{eqnarray}
\label{deriv} \frac{\partial\tilde{\mathcal
I}_1(A_0,h,\tilde{h})}{\partial \tilde{h}}=
-i\int_{-\infty}^{\infty}d\lambda\;
\!\int_{-i\infty}^{i\infty}d{\tilde \lambda}\;
\;\exp\left\{-\frac{A_0}{2}\;(\lambda^2-{\tilde \lambda}^2)
+ih\lambda-i\tilde{h}\tilde{\lambda }\right\}\nonumber\\
=\sqrt{\frac{2\pi}{A_0}}\;e^{-h^2/(2A_0)}\;\sqrt{\frac{2\pi}{A_0}}\;e^{\tilde{h}^2/(2A_0)}=\frac{2\pi}{A_0}
\exp\left\{-\frac{h^2-\tilde{h}^2}{2A_0}\right\}.
\end{eqnarray}
Then the function $\tilde{\mathcal I}_1$ itself will have the form
\begin{eqnarray}
\label{form} \tilde{\mathcal
I}_1(A_0,h,\tilde{h})=\frac{2\pi}{A_0}\;e^{-h^2/(2A_0)}\left(\int_0^{\tilde{h}}d\tilde{h}_1e^{\;\tilde{h}_1^2/(2A_0)}+C\right),
\end{eqnarray}
with $C=(A_0/2\pi)\exp(h^2/2A_0)\tilde{\mathcal I}_1(A_0,h,0)$. On
the other hand we have that
\begin{eqnarray}
\label{zero} \tilde{\mathcal I}_1(A_0,h,0)=
\int_{-\infty}^{\infty}d\lambda\;
\text{\large{P}}\!\int_{-i\infty}^{i\infty}\frac{d{\tilde
\lambda}}{{\tilde \lambda}}\;
\;\exp\left\{-\frac{A_0}{2}\;(\lambda^2-{\tilde \lambda}^2)
+ih\lambda\right\}=0
\end{eqnarray}
suggesting $C=0$. Substituting Eq. (\ref{form}) into Eq.
(\ref{Int1}) we obtain
\begin{eqnarray}
\label{finalform} {\mathcal I}_1(A_0,h,{\tilde
h})=\pi\sqrt{\frac{2\pi}{A_0}}\exp\left(-\frac{h^2}{2A_0}\right)
\left[i+\text{erfi}\left(\frac{\tilde{h}}{\sqrt{2A_0}}\right)\right].
\end{eqnarray}
\subsection{Calculation of ${\mathcal I}_2$}
Introducing new variable, $\lambda^{\prime}=-i\tilde\lambda$, we
rewrite integral $ {\mathcal I}_2(h,\tilde{h})$ as
\begin{eqnarray}
\label{I2}
 {\mathcal I}_2(h,\tilde{h})= i\int_{-\infty}^{\infty}d\lambda\int_{-\infty}^{\infty}d\lambda^{\prime}\;\frac{1}
{\lambda-i
\lambda^{\prime}}\;\exp\left\{-\frac{A_0}{2}\left(\lambda-\frac{ih}{A_0}\right)^2-\frac{A_0}{2}\left(\lambda^{\prime}
-\frac{\tilde{h}}{A_0}\right)^2-\frac{h^2-\tilde{h}^2}{2A_0}\right\}.
%i(h\lambda-i\tilde{h}\lambda^{\prime})\right\}.
\end{eqnarray}
As the next step we shift variables $\lambda$ and
$\lambda^{\prime}$ by $ih/A_0$ and $\tilde{h}/A_0$ respectively.
Then Eq. (\ref{I2}) will acquire the form
\begin{eqnarray}
\label{shift}
 {\mathcal I}_2(h,\tilde{h})=ie^{-\frac{h^2-\tilde{h}^2}{2A_0}}\int_{-\infty}^{\infty}d\lambda\int_{-\infty}^{\infty}d\lambda^{\prime}\;
 \frac{e^{-\frac{A_0}{2}\left(\lambda^2+{\lambda^\prime}^2\right)}}{\lambda+\frac{ih}{A_0}-i\left(\lambda^\prime+\frac{\tilde{h}}{A_0}\right)}.
\end{eqnarray}
It is convenient to evaluate integral Eq. (\ref{shift}) after
switching to polar coordinates,
$\lambda-i\lambda^{\prime}=ue^{i\theta}$:
%To evaluate Eq. (\ref{shift}) we pass to polar coordinates,
%$\lambda-i\lambda^{\prime}=ue^{i\theta}$,
\begin{eqnarray}
\label{polar} {\mathcal
I}_2(h,\tilde{h})&=&ie^{-\frac{h^2-\tilde{h}^2}{2A_0}}
\int_0^{\infty}u\;du\int_0^{2\pi}d\theta\;
\frac{e^{-\frac{A_0}{2}u^2}}{\frac{ih}{A_0}-\frac{i\tilde{h}}{A_0}+ue^{i\theta}},
\end{eqnarray}
and integrate first over $\theta$ and only then over $u$. Result
reads
\begin{eqnarray}
\label{polarInt} {\mathcal
I}_2(h,\tilde{h})&=&ie^{-\frac{h^2-\tilde{h}^2}{2A_0}}\frac{2\pi}{i\left(\frac{h}{A_0}-\frac{\tilde{h}}{A_0}\right)}\left(1-\delta_{h\tilde{h}}\right)
\int_0^{\infty}du\;ue^{-\frac{A_0}{2}u^2}\nonumber\\
&=&\frac{2\pi}{h-\tilde{h}}\;\left(1-\delta_{h\tilde{h}}\right)e^{-\frac{h^2-\tilde{h}^2}{2A_0}}.
\end{eqnarray}
%%%%%%%%%%%%%%%%%%%%%%%%%%%%%%%%%%%%%%%%%
%%%%%%%%%%%%%%%%%%%%%%%%%%%%%%%%%%%%%%%%%%%%%%%%%%%
%%%%%%%%%%%%%%%%%%%%%%%%%%%%%%%%%%%%%%%%%%%%%%%%%%%
\end{widetext}


\begin{thebibliography}{99}
\bibitem{footnote}
It is useful to note that formal sums of formal products $\psi \otimes {\bar{%
\psi}}$, where $\psi \in U(n,1|n,1)$ and ${\bar{\psi}}\in U(n,1|n,1)$ are
supervectors, constitute a vector space. This vector space is defined, up to
isomorphism, by the condition that every antisymmetric, bilinear map $%
f:U(n,1|n,1)\times {\bar{U}}(n,1|n,1)\rightarrow {\mathbb{G}}$ determines a
unique linear map $g:U(n,1|n,1)\otimes {\bar{U}}(n,1|n,1)\rightarrow {%
\mathbb{G}}$ with $f(\psi ,{\bar{\psi}})=g(\psi \otimes {\bar{\psi}})$. This
implies that if we consider a map, $F:{\mathbb{H}}_{n}\rightarrow {\mathbb{G}%
}$, then the integrals in Eq.~(\ref{formula}) are well defined. Both
integrals in Eq.~(\ref{formula}) throughout this article are assumed to be
convergent. The question of classification of maps $F$, for which this
property holds (addressed in Ref.~\onlinecite{Littelmann}), is beyond the
scope of the current work.
\end{thebibliography}
\end{document}